\documentclass[conference]{IEEEtran}
\IEEEoverridecommandlockouts
\usepackage{cite}
\usepackage{amsmath,amssymb,amsfonts}
\usepackage{algorithmic}
\usepackage{graphicx}
\usepackage{textcomp}
\usepackage{xcolor}

\usepackage{bm}    
\usepackage{amsthm}
\newtheorem{thm}{\hspace{1em}Theorem}

\usepackage{xpatch}
\makeatletter
\xpatchcmd{\@thm}{\thm@headpunct{.}}{\thm@headpunct{:}}{}{}
\makeatother
\newtheorem{lemma}{\hspace{1em}Lemma}

\usepackage{algorithm,algorithmic}
\makeatletter
\newcommand{\removelatexerror}{\let\@latex@error\@gobble}
\makeatother

\usepackage{url}  

\usepackage{subfigure}   

\usepackage[switch]{lineno}

\def\BibTeX{{\rm B\kern-.05em{\sc i\kern-.025em b}\kern-.08em
		T\kern-.1667em\lower.7ex\hbox{E}\kern-.125emX}}
\usepackage{cite}

\usepackage{bbding}

\usepackage{fancyhdr}

\begin{document}
	
	\title{CGN: A Capacity-Guaranteed Network Architecture for Future Ultra-Dense Wireless Systems\\	}
	
	\author{\IEEEauthorblockN{Chaowen Deng\IEEEauthorrefmark{2}, Lu Yang\IEEEauthorrefmark{3} \Envelope, Hao Wu\IEEEauthorrefmark{2}, Dmitry Zaporozhets\IEEEauthorrefmark{4}, Miaomiao Dong\IEEEauthorrefmark{3} and Bo Bai\IEEEauthorrefmark{3}}
		\IEEEauthorblockA{\IEEEauthorrefmark{2}Department of Mathematical Sciences, Tsinghua University, Beijing, China}
		\IEEEauthorblockA{\IEEEauthorrefmark{3}Theory Lab, Central Research Institute, 2012 Labs, Huawei Technology Co. Ltd.}
		\IEEEauthorblockA{\IEEEauthorrefmark{4}Steklov Mathematical Institute of Russian Academy of Sciences, St. Petersburg, Russia}
		\IEEEauthorblockA{Email: dcw21@mails.tsinghua.edu.cn, yanglu87@huawei.com,  hwu@tsinghua.edu.cn, \\zap1979@gmail.com, dong.miaomiao@huawei.com, baibo8@huawei.com}
	}

	\maketitle
	
	\thispagestyle{fancy}
	\fancyhead{}
	\lhead{}
	\lfoot{\small\copyright~2022 IEEE. Personal use of this material is permitted.  Permission from IEEE must be obtained for all other uses, in any current or future media, including reprinting/republishing this material for advertising or promotional purposes, creating new collective works, for resale or redistribution to servers or lists, or reuse of any copyrighted component of this work in other works.}
	\cfoot{}
	\rfoot{}

	\begin{abstract}
		The sixth generation (6G) era is envisioned to be a fully intelligent and autonomous era, with physical and digital lifestyles merged together. Future wireless network architectures should provide a solid support for such new lifestyles. A key problem thus arises that what kind of network architectures are suitable for 6G. In this paper, we propose a capacity-guaranteed network (CGN) architecture, which provides high capacity for wireless devices densely distributed everywhere, and ensures a superior scalability with low signaling overhead and computation complexity simultaneously. Our theorem proves that the essence of a CGN architecture is to decompose the whole network into non-overlapping clusters with equal cluster sum capacity. Simulation results reveal that in terms of the minimum cluster sum capacity, the proposed CGN can achieve at least 30\% performance gain compared with existing base station clustering (BS-clustering) architectures. In addition, our theorem is sufficiently general and can be applied for networks with different distributions of BSs and users.
	\end{abstract}
	
	
	\section{Introduction}
	The 6G era is envisioned to revolutionize wireless services via the Internet of Things (IoT) towards the future of fully intelligent and autonomous systems \cite{9509294}. Unquestionably, with the future commercialization of 6G, our lifestyles will be greatly changed. More and more intelligent and digital services will emerge, such as robotic surgery, human-machine communications, blockchain-based production, satellite-unmanned aerial vehicles communications, etc. These services will accelerate the fusion of human, physical and digital worlds, but will also induce the explosive increase of the wireless data traffic. These phenomena will bring about extremely high pressure on the network architecture side, since it should not only satisfy the ever-growing requirement of the wireless data traffic, but also guarantee high-quality services for all the wireless devices densely located everywhere. Then an important problem arises: whether existing network architectures are still suitable for the 6G era?
	
	The cellular network is the most classic architecture \cite{2}. It is composed of cells, where each cell corresponds to the coverage area of a single base station (BS). This architecture is designed based on the BS-side information, and thus can be regarded as a BS-centric architecture \cite{8422218}. The main bottleneck of the cellular architecture is the cell-edge problem, which refers to that users located at the cell edges suffer from serious inter-cell interference \cite{5594708}. Another BS-centric architecture is the BS-clustering architecture \cite{8422218}, where several closely-located BSs form a cluster to serve users jointly, through techniques like coordinated multipoint transmissions and receptions. However, there are still users suffering from the cluster-edge problem.
	
	In order to eliminate the cell/cluster-edge problems in BS-centric architectures, the user-centric manner was proposed in \cite{6941550, 8387197}. Each user autonomously chooses nearby BSs to form its own cell, called a virtual cell. These virtual cells are overlapped with each other with a high probability. One common way to deal with the overlapping cells is to merge them. Thus, large clusters exist in such user-centric architectures, where the signaling overhead and the measurement complexity are very high.
	
	Another existing architecture is the cell-free massive multiple-input multiple-output (MIMO) \cite{7827017}, where all the BSs are interconnected through backhaul links to serve all the users jointly with one or several central processing units (CPUs). There are no cells and thus no cell boundaries. The main bottleneck of this architecture is its unscalability \cite{Matthaiou2021TheRT}, since the induced signaling overhead over the backhaul links and the computation complexity at the CPUs expand much more faster than the increase of BSs and users. Actually, the number of coordinated BSs is always limited in real-world deployments. As analyzed above, existing architectures are not suitable for future ultra-dense wireless networks.
	
\begin{figure}[htbp]
		\centerline{\includegraphics[scale=0.076]{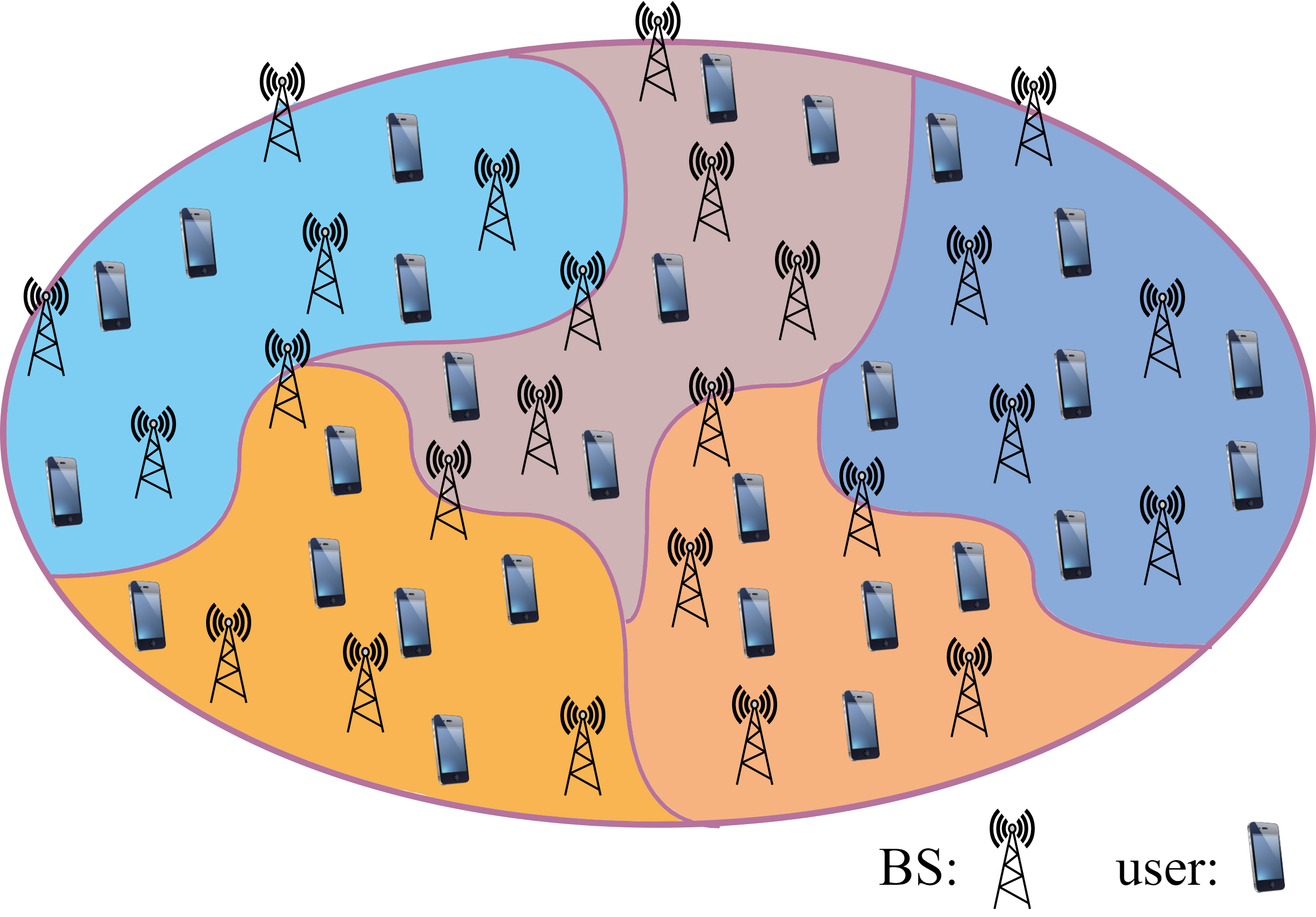}}
		\caption{A CGN architecture diagram. An ultra-dense wireless system is decomposed into non-overlapping clusters according to the principle of maximizing the minimum cluster sum capacity.}
		\label{CAG}
	\end{figure}

	In this paper, we start from an analytical point of view and propose a brand-new architecture, called the $C$apacity-$G$uaranteed $N$etwork (CGN) architecture. CGN architectures can guarantee the users densely-distributed everywhere with a sufficiently high quality of service (QoS), and ensure the low signaling overhead and computation complexity simultaneously. On one hand, to realize the high QoS everywhere and mitigate the cell-edge problem, we adopt the design principle of maximizing the minimum cluster sum capacity, which will be elaborated later in Section \ref{PF}. On the other hand, to ensure that our CGN architecture is scalable with low signaling overhead, we decompose the whole network into non-overlapping clusters with each cluster operating independently \cite{yang}, where the number of coordinated BSs is limited, and thus the signaling overhead and the computation complexity are constrained by the cluster size. Therefore, the CGN architecture can realize both high capacity and high scalability simultaneously, and thus is more suitable for future ultra-dense wireless systems. We illustrate a CGN diagram in Fig. \ref{CAG}.
	
	\section{System Model And Problem Formulation}
	\subsection{System Model}
	Consider an ultra-dense wireless network consisting of $M$ single-antenna BSs\footnote{For distributed antenna systems, we consider $M$ single-antenna access points (APs), instead of BSs.} and $K$ single-antenna users\cite{7827017, 7841875}. To capture the ultra-dense property of 6G, we assume both $K$ and $M$ approaching positive infinity \cite{yang}. Denote the set of BSs as $\mathcal{B} = \{b_{1},..., b_{M}\}$, and the set of users as $\mathcal{U} = \{u_{1},..., u_{K}\}$. Assume the network is decomposed into $L$ non-overlapping clusters, with each cluster operating independently. The set of network nodes (BSs and users) of the $l$-th ($l=1,2,...,L$) cluster is denoted by $\mathcal{S}_l$, and $\Theta = \{\mathcal{S}_{1}, \mathcal{S}_{2}, ..., \mathcal{S}_{L}\}$ forms a partition of the set $\mathcal{B} \cup \mathcal{U}$ . In $\mathcal{S}_{l}$, the number of users is $K_{l}$, and the number of BSs is $M_{l}$. Thus, $K = \sum_{l=1}^{L}K_{l}$ and $M = \sum_{l=1}^{L}M_{l}$.
	
	The channel gain between the user $u_k \in \mathcal{U}$ and the BS $b_m \in \mathcal{B}$ is denoted by $h_{mk}$, which is modeled as:
	\begin{equation}
		\label{h}
		h_{mk} = \beta_{mk}\gamma_{mk},
	\end{equation}
	where $\gamma_{mk}$ is the small-scale fading, following the complex Gaussian distribution $\mathcal{CN}(0,1)$; $\beta_{mk}$ represents the large-scale fading, defined as
	\begin{equation}
		\label{beta}
		\begin{split}
			\beta_{mk} = \left\{
			\begin{array}{ll}
				d_{mk}^{-\frac{\alpha}{2}}, &d_{mk} > d_{0},\\
				d_{0}^{-\frac{\alpha}{2}}, &d_{mk} \le d_{0},
			\end{array}
			\right.
		\end{split}
	\end{equation}
	where $d_{mk}$ is the Euclidean distance between $b_m$ and $u_k$, $\alpha$ is the path-loss factor, and $d_0$ is a distance threshold, which is used to avoid $\beta_{mk}$ approaching infinity when $b_m$ and $u_k$ located extremely close. Assume each user transmits with power $P$, and the background noise power is $N_{0}$.
	
	\subsection{Problem Formulation}\label{PF}
	Our objective is to design a novel wireless network architecture which not only ensures the scalability of the system, but also guarantees high communication quality everywhere. To realize the high scalability with low signaling and computational overhead, we decompose the whole system into $L$ ($L>1$) non-overlapping clusters. To guarantee the high communication quality, we choose the principle of maximizing the minimum cluster sum capacity. Such an objective can be formulated as
	\begin{equation*}
		\mathcal{P}1: ~~
		\label{max-min optimal problem}
		\max \limits_{\Theta}   \min \limits_{l\in \{1,2...L\}} C_{l},
	\end{equation*}
	where $C_{l}$ is the uplink sum capacity of the $l$-th cluster. 
	
	In order to solve $\mathcal{P}$1, the expression of $C_{l}$ should be determined at first. In this paper, we focus on the uplink case. The uplink sum capacity of the $l$-th cluster can be determined according to \cite{2} as
	\begin{equation}
		\label{second sum capacity}
		C_{l} \!= \!\mathbb{E} \left[ \log \det\! \left( \mathbf{I}_{M_{l}}\!\!+\!\!P \left( N_{0}\mathbf{I}_{M_{l}}\!\!+\!\!P\mathbf{\Pi}_{l}\mathbf{\Pi}_{l}^{H} \right)\!\! ^{-1}\!\mathbf{H}_{l}\mathbf{H}_{l}^{H} \right) \right],
	\end{equation}
	where $\mathbf{H}_{l} \in \mathbb{C}^{M_{l} \times K_{l}}$ and $\bm{\Pi}_{l} \in \mathbb{C}^{M_{l} \times (K-K_{l})}$ denote the channel gain matrix and the interference matrix of the $l$-th cluster, respectively. Specifically, the $(m,l)$-th entry of $\mathbf{H}_{l}$ can be determined according to (\ref{h}), with $b_{m} \in \mathcal{S}_{l}$, and $u_{k} \in \mathcal{S}_{l}$. The entries of $\bm{\Pi}_{l}$ can be determined based on (\ref{h}) as well, but with $b_{m} \in \mathcal{S}_{l}$ and $u_{k} \in \mathcal{U} \backslash \mathcal{S}_l$\footnote{$\mathcal{U} \backslash \mathcal{S}_l$ denotes the set of users in $\mathcal{U}$ but not in $\mathcal{S}_l$.}. The term $P \left( N_{0}\mathbf{I}_{M_{l}}+P\mathbf{\Pi}_{l}\mathbf{\Pi}_{l}^{H} \right) ^{-1}\mathbf{H}_{l}\mathbf{H}_{l}^{H}$ represents the signal to the interference plus noise ratio (SINR) of the $l$-th cluster. The asymptotic property of $\mathbf{\Pi}_{l}\mathbf{\Pi}_{l}^{H} = (\pi_{ij})_{M_{l} \times M_{l}}$ is shown in Lemma \ref{theorem 1}.
	
	\begin{lemma}
		\label{theorem 1}As $K - K_{l} \to \infty$,   $\mathbf{\Pi}_{l}\mathbf{\Pi}_{l}^{H} \in \mathbb{C}^{M_{l} \times M_{l}}$ converges to a diagonal matrix, which is
		\begin{equation}
			\mathbf{\Pi}_{l}\mathbf{\Pi}_{l}^{H} = {\rm diag} (\pi_{11},...,\pi_{mm},...,\pi_{M_{l}M_{l}}),
		\end{equation}
		whose $m$-th diagonal entry is
		\begin{equation}
			\label{diag2}
			\pi_{mm} = \sum_{u_{k} \in \mathcal{U} \backslash \mathcal{S}_l}\beta_{mk}^2,
		\end{equation}
	\end{lemma}
	\begin{IEEEproof}[$Proof$]First, we prove the off-diagonal entries of $\mathbf{\Pi}_{l}\mathbf{\Pi}_{l}^{H}$ approaching zero as $K - K_{l} \rightarrow \infty$. Based on the definition of $\mathbf{\Pi}_l$, we can obtain the $(i,j)$-th entry $(i \neq j)$ of $\mathbf{\Pi}_{l}\mathbf{\Pi}_{l}^{H}$ as
		\begin{equation}
			\label{inter mat}
			\pi_{ij} = \sum_{u_{k} \in \mathcal{U} \backslash \mathcal{S}_l} \left( \beta_{ik}\gamma_{ik} \right) \left( \overline{\beta_{jk}}\overline{\gamma_{jk}} \right).
		\end{equation}
		Since $\gamma_{ik}$ is a $\mathcal{CN}(0,1)$ random variable (RV), we let $\gamma_{ik}=a_{ik} + \mathbf{i}b_{ik}$. Both $a_{ik}$ and $b_{ik}$ are identically distributed real Gaussian RVs following $\mathcal{N}(0,\frac{1}{2})$. Then, (\ref{inter mat}) can be written as
		\begin{equation}
			\label{inter mat 2}
			\pi_{ij} =\!\!\!\! \sum_{u_{k} \in \mathcal{U} \backslash \mathcal{S}_l} \!\!\!\! \beta_{ik}\beta_{jk} \left[ a_{ik}a_{jk} \!\! + \! b_{ik}b_{jk} +\textbf{i} \left( a_{jk}b_{ik} \!\! -\!\! a_{ik}b_{jk} \right) \right].
		\end{equation}
		Note that the product of two independent real Gaussian RVs is also a real Gaussian RV. According to Chebyshev's theorem\footnote{Chebyshev's theorem: If $\{X_{i}\}, i=1,...,n$, are independent RVs with $E\{X_{i}\}=\mu_{i}$ and $Var\{X_{i}\} \le c < +\infty$, then \\ \centerline{$\frac{1}{n} \sum_{i=1}^{n}X_{i} - \frac{1}{n} \sum_{i=1}^{n}\mu_{i} \mathop{\to}\limits^{p} 0$.} }, (\ref{inter mat 2}) approaches zero as $K - K_{l} \to \infty$, so the off-diagonal entries of $\mathbf{\Pi}_{l}\mathbf{\Pi}_{l}^{H}$ approach zero. 
		
		Next, let's study the diagonal entries of $\mathbf{\Pi}_{l}\mathbf{\Pi}_{l}^{H}$. Similar to the above procedures, the $m$-th diagonal entry can be derived as
		\begin{equation}
			\label{diag}
			\pi_{mm} = \sum_{u_{k} \in \mathcal{U} \backslash \mathcal{S}_l}\beta_{mk}^2(a_{mk}^{2}+b_{mk}^{2}),
		\end{equation}
		where $(\sqrt{2}a_{mk})^{2}+(\sqrt{2}b_{mk})^{2} \sim \chi^{2}(2)$ (chi-square distribution). According to Chebyshev's theorem, we can obtain (\ref{diag2}) from (\ref{diag}). 
		
		The proof is completed.
	\end{IEEEproof}
	\medskip
	Based on Lemma \ref{theorem 1}, (\ref{second sum capacity}) can be written as
	\begin{equation}
		\label{third sum capacity}
		C_{l} = \mathbb{E} \left[ \log \det \left( \mathbf{I}_{M_{l}}+P\mathbf{D}_{l}^{-1}\mathbf{H}_{l}\mathbf{H}_{l}^{H}\right) \right],
	\end{equation}
	where $\mathbf{D}_{l} = N_{0}\mathbf{I}_{M_{l}}+P\mathbf{\Pi}_{l}\mathbf{\Pi}_{l}^{H}$ is a diagonal matrix. We will use (\ref{third sum capacity}) for cluster sum capacity calculation in section \ref{AVAPE}.

	\section{CGN Architecture Theorem}\label{NNAAA}
	In this section, we will elaborate the core mathematical principles of CGN architectures. The theorem which gives the solution of $\mathcal{P}$1 and reveals the essence of future network architecture designs is given below. 
	
	\begin{thm}\label{thm} In ultra-dense scenario, the optimal network decomposition can be achieved only if all the clusters have the same cluster sum capacity, which is: 
		\begin{equation}
			\label{Cequal}
			C_{1} = C_{2} = ... = C_{L},
		\end{equation}
	\end{thm}
	\begin{IEEEproof}[$Proof$]We utilize the method of proof by contradiction. With a slight abuse of notation, here we denote the optimal network decomposition as $\Theta$. It means the minimum cluster sum capacity in $\Theta$ is the maximum one among all the network decompositions. 
		
		If (\ref{Cequal}) does not hold in $\Theta$, there exists a minimum cluster sum capacity in $\Theta$, denoted as $C_{min}$. Let $\Theta_{min}$ be the set of clusters whose sum capacity equal to $C_{min}$. Let $|\Theta_{min}|$ denote the cardinality of $\Theta_{min}$.
		
		$Case\hspace{0.5em} 1$: $|\Theta_{min}| = 1$. It means there is only one cluster with the capacity of $C_{min}$. Assume this cluster is $\mathcal{S}_{i}$. Then we have $\Theta_{min} = \big\{ \mathcal{S}_{i} \big\}$, and $C_{i} = C_{min} < C_{l},(l\neq i).$
		Denote $C_{j} = \min\limits_{l \neq i} \{ C_{l} \}$, and thus
		\begin{equation}
			\label{varepsilon}
			\varepsilon = C_{j} - C_{i} > 0.
		\end{equation}
		Let's consider a new decomposition generated from $\Theta$ as $\Theta^{*}$, where $C_{i}$ increases by an increment $\frac{\varepsilon}{3}$ and the reduction of the cluster sum capacity of other clusters will not exceed $\frac{\varepsilon}{3}$:
		\begin{equation}
			\begin{cases}
				\label{update network_1}
				C_{i}^{*} = C_{i} + \frac{\varepsilon}{3},\\
				C_{l}^{*} \geq C_{l} - \frac{\varepsilon}{3}, (l \neq i).
			\end{cases}
		\end{equation}
		Thus $C_{i}^{*}$ is the minimum cluster sum capacity in $\Theta^{*}$. However,  $C_{i}^{*} > C_{i}$, which means $\Theta$ is not the optimal decomposition with the maximized $C_{min}$. This leads to a contradiction with the original assumption that $\Theta$ is the optimal decomposition.
		
		$Case \hspace{0.5em} 2$: $|\Theta_{min}| = n$ $(2 \leq n < L)$\footnote{If $n=L$, the equation $C_{1} = C_{2} = ... = C_{L}$ holds directly.}. It means there are multiple clusters with the capacity of $C_{min}$. Then we have
		\begin{equation}
			\begin{cases}
				\label{n}
				C_{i} = C_{min}, \, \forall \mathcal{S}_{i} \in \Theta_{min}\\
				C_{i} < C_{l}, \:\:\:\:\:\: \forall \mathcal{S}_{l} \in \Theta \backslash \Theta_{min} .
			\end{cases}
		\end{equation}
		We randomly select a cluster from $\Theta_{min}$, might as well take $\mathcal{S}_{i}$. Denote $C_{j} = \min \limits_{\mathcal{S}_{l} \in \Theta \backslash \Theta_{min}} \{C_{l}\}$, and thus
		\begin{equation}
			\label{varepsilon_2}
			\varepsilon = C_{j} - C_{i} > 0.
		\end{equation}
		Using the same argument as $Case \hspace{0.5em} 1$, we have the following relationships in $\Theta^*$:
		\begin{equation}
			\begin{cases}
				\label{update network_2}
				C_{i}^{*} = C_{i} + \frac{\varepsilon}{3},\\
				C_{l}^{*} \geq C_{l} - \frac{\varepsilon}{3}, \forall \mathcal{S}_{l} \in \Theta \backslash \Theta_{min}.
			\end{cases}
		\end{equation}
		For the cluster with $C_{min}$ in $\Theta^{*}$, there exist three possibilities:
		\begin{itemize}
			\item[ (i)] There is a single cluster belonging to $\Theta_{min} \backslash \{\mathcal{S}_{i}\}$ with its capacity dropped the most, which means $|\Theta^*_{min}|=1$. We can use the same procedures to get the contradiction as shown in $Case \hspace{0.5em} 1$.
			\item[ (ii)] There are more than one clusters in the set $ \Theta_{min} \backslash \{\mathcal{S}_{i}\}$ with the capacity decreased by a same value. Then the situation becomes $1<|\Theta^*_{min}| \leq n-1$. We can repeat the procedures from the initialization of $Case \hspace{0.5em} 2$ until the contradiction happens.
			\item[ (iii)] None of the clusters in $\Theta_{min} \backslash \{\mathcal{S}_{i}\} $ with the capacity changed. It means there are cluster(s) in $\Theta \backslash \Theta_{min}$ with capacity decreased, and thus $|\Theta^{*}_{min}|=n-1$. We can then repeat the procedures from the initialization of $Case \hspace{0.5em} 2$ until the contradiction happens.
		\end{itemize}
		
		The proof is completed.
	\end{IEEEproof}
	
	Theorem \ref{thm} gives the optimal solution of $\mathcal{P}$1, and reveals that the optimal network architecture must satisfy the condition of equal cluster sum capacity, i.e., $C_{1} = C_{2} = ... = C_{L}$. This is the essence of our CGN architectures. As such, our CGN can be regarded as a capacity-centric architecture, which is designed based on the information from the BS side and the user side jointly. As a contrast, conventional cellular and BS-clustering architectures are BS-centric only, missing the important information from the user side.
	
	A key point we want to emphasize is that Theorem \ref{thm} does not impose any restrictions on the network area and the distributions of BSs and users. The only constraint is the ultra-dense scenario, which is obviously satisfied in the future 6G era aiming to connect everything. Therefore, we claim that Theorem \ref{thm} and our CGN architecture have superior generality and a wide range of applications. They are feasible for various wireless networks with different distributions of BSs and users. Besides, the number of clusters $L$ is adaptive to concrete network requirements. If the BS coordination can not be performed in a certain network, we can set $L=M$, where the number of clusters equals to the number of BSs. Then an extreme case of CGN arises, which is equivalent to the cellular architecture.
	
	\section{Architecture Visualization And Performance Evaluation}\label{AVAPE}
	In this section, we will visualize our proposed CGN architecture through a heuristic algorithm, and show the performance comparison across CGN, BS-clustering and cellular architectures through simulations.
	
	\subsection{Heuristic CGN Architecture Algorithm}
	Denote the cluster with the maximum cluster sum capacity $C_{max}$ as $\mathcal{S}_{max}$ and the cluster with the minimum cluster sum capacity $ C_{min}$ as $\mathcal{S}_{min}$. Based on our Theorem \ref{thm}, we propose a heuristic algorithm with the objective to realize a network decomposition with $C_{max}\!-\!C_{min}\!\! \!\rightarrow \!0$. Details are shown in Algorithm \ref{algorithm}, where $center$ denotes an $L \times 2$ matrix with each row vector representing the centroid coordinates of each cluster, $idx$ denotes a $(K+M) \times 1$ vector with each entry representing the index of the clusters that a BS/user belongs to, $center_{\mathcal{S}_{max}}$ denotes the centroid coordinates of $\mathcal{S}_{max}$. Define $d(p,\mathcal{S}) = \min \limits_{q\in \mathcal{S}} d(p,q)$ as the minimum Euclidean distance between the point $p$ and  the point $q \in \mathcal{S}$.
	\begin{figure}[h]
		\vspace{-2em}
		\renewcommand{\algorithmicrequire}{\textbf{Input:}}
		\renewcommand{\algorithmicensure}{\textbf{Output:}}
		\removelatexerror
		\begin{algorithm}[H]
			\caption{Heuristic CGN Architecture Algorithm}
			\label{algorithm}
			\begin{algorithmic}[1]
				\REQUIRE Coordinates of users and BSs, $K$, $M$, $L$, and $\delta$.         
				\ENSURE $center$, $idx$ and $C_l (l = 1,2...,L)$.  
				\STATE{\textbf{Initialization:} Decompose the whole network into $L$ clusters based on $k$-means++ algorithm.} \label{thmstep1}
				\STATE {Compute $C_l (l = 1,2...,L)$ based on (\ref{third sum capacity}) and find $C_{max}$ and $C_{min}$.}
				\WHILE{$C_{max} - C_{min}$ $>$ $\delta$}
				
				\FOR{$p \in (\mathcal{B} \cup \mathcal{U}) \backslash \mathcal{S}_{min}$}
				\IF{$p = \arg \min \limits_{p^{*} \in (\mathcal{B} \cup \mathcal{U})  \backslash \mathcal{S}_{min}} d (p^{*},\mathcal{S}_{min}$)}
				\STATE {Let $p \in \mathcal{S}_{min}$.}
				\ENDIF
				\ENDFOR
				
				\FOR{$q \in \mathcal{S}_{max}$}
				\STATE $center_{\mathcal{S}_{max}} \leftarrow$ the centroid coordinate of $\mathcal{S}_{max}$
				\IF{$q = \arg \max \limits_{q^{*} \in \mathcal{S}_{max}} d (q^{*},center_{\mathcal{S}_{max}})$}
				\STATE Find $\mathcal{S}_{q} = \arg \min \limits_{\mathcal{S} \neq \mathcal{S}_{max}} d (q,\mathcal{S})$.
				\STATE Let $q \in \mathcal{S}_{q}$.
				\ENDIF
				\ENDFOR
				\STATE {Update $center$ and $idx$.}
				\STATE {Update $C_{max}$ and $C_{min}$.}
				\ENDWHILE
				\RETURN $center$, $idx$ and $C_l (l = 1,2...,L)$
			\end{algorithmic}
		\end{algorithm}
	\end{figure}
	
	\begin{figure*}[ht]
		\centering
		\hspace{-0.05\linewidth}
		\subfigure[]{\label{au}
			\includegraphics[width=0.42\linewidth]{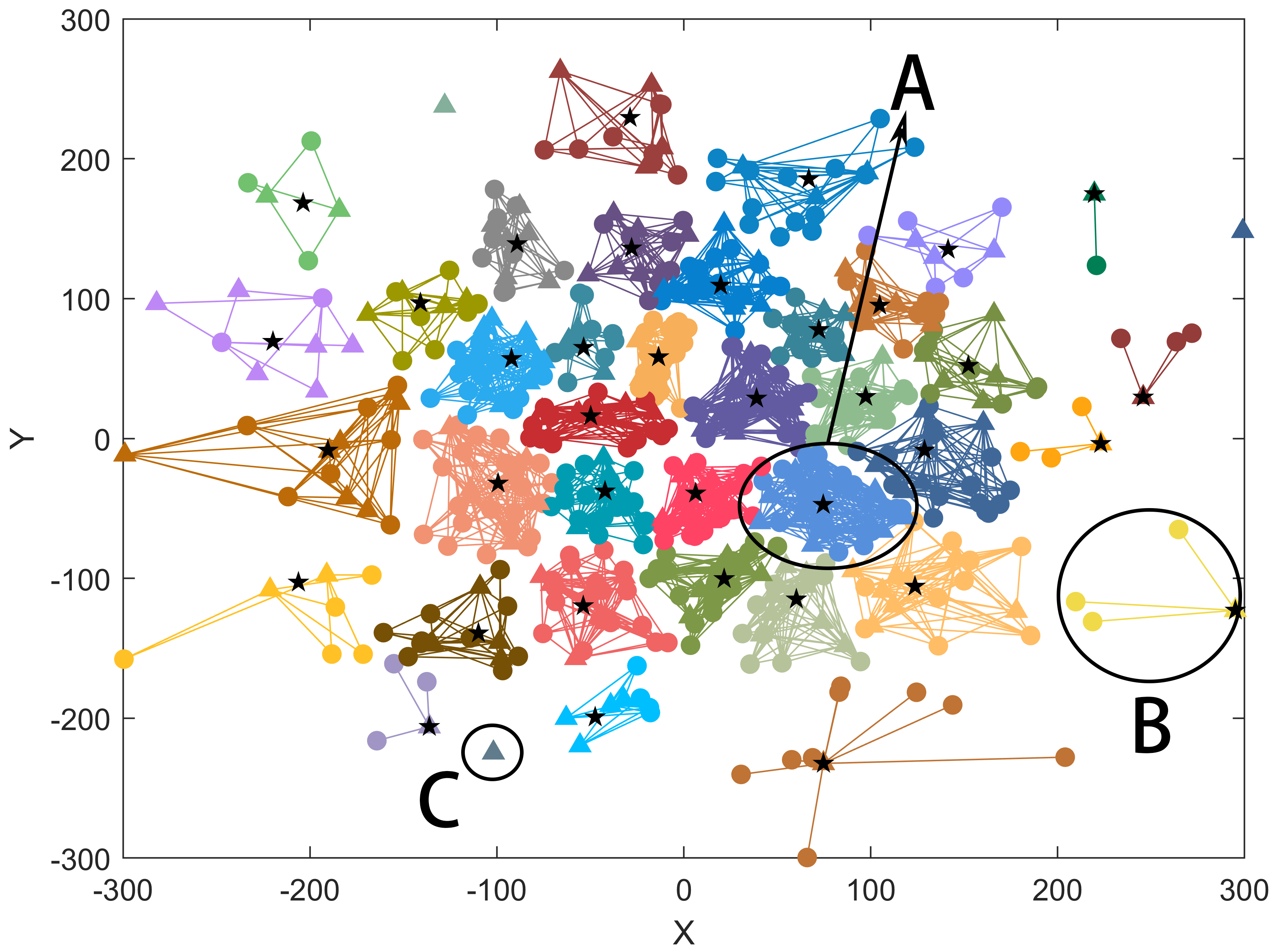}}
		\hspace{0.05\linewidth}
		\subfigure[]{\label{bu}
			\includegraphics[width=0.42\linewidth]{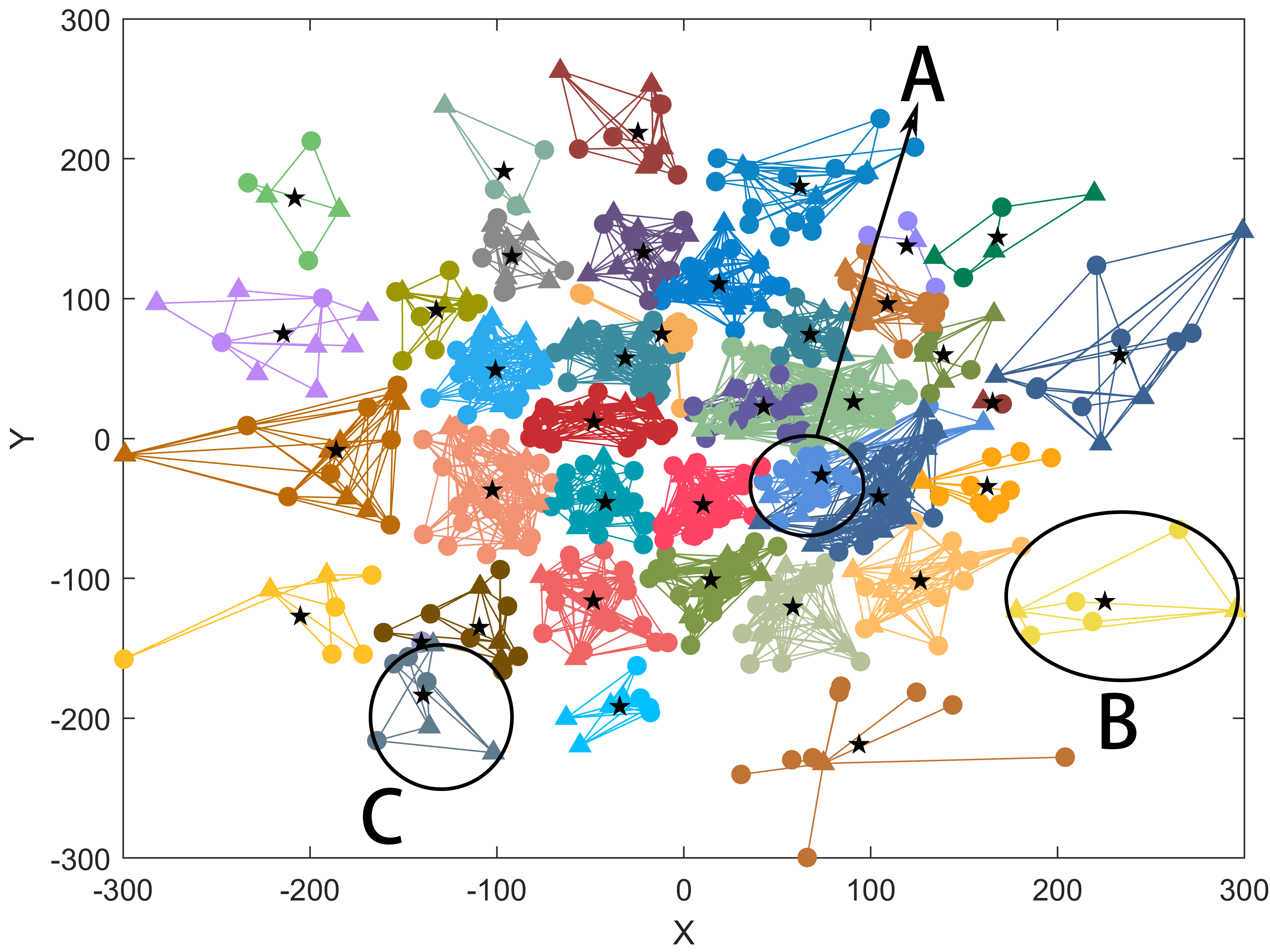}}
		\caption{Visualization of BS-clustering and CGN architectures, with $K>M$. $K = 400, M=200$. Both BSs and users are distributed according to GD. Triangles represent BSs, circles represent users, and stars represent the centroids of clusters. (a) BS-clustering. (b) CGN.}
		\label{The results of simulation of u}
	\end{figure*}
	\begin{figure*}[ht]
		\centering
		\hspace{-0.05\linewidth}
		\subfigure[]{\label{ab}
			\includegraphics[width=0.42\linewidth]{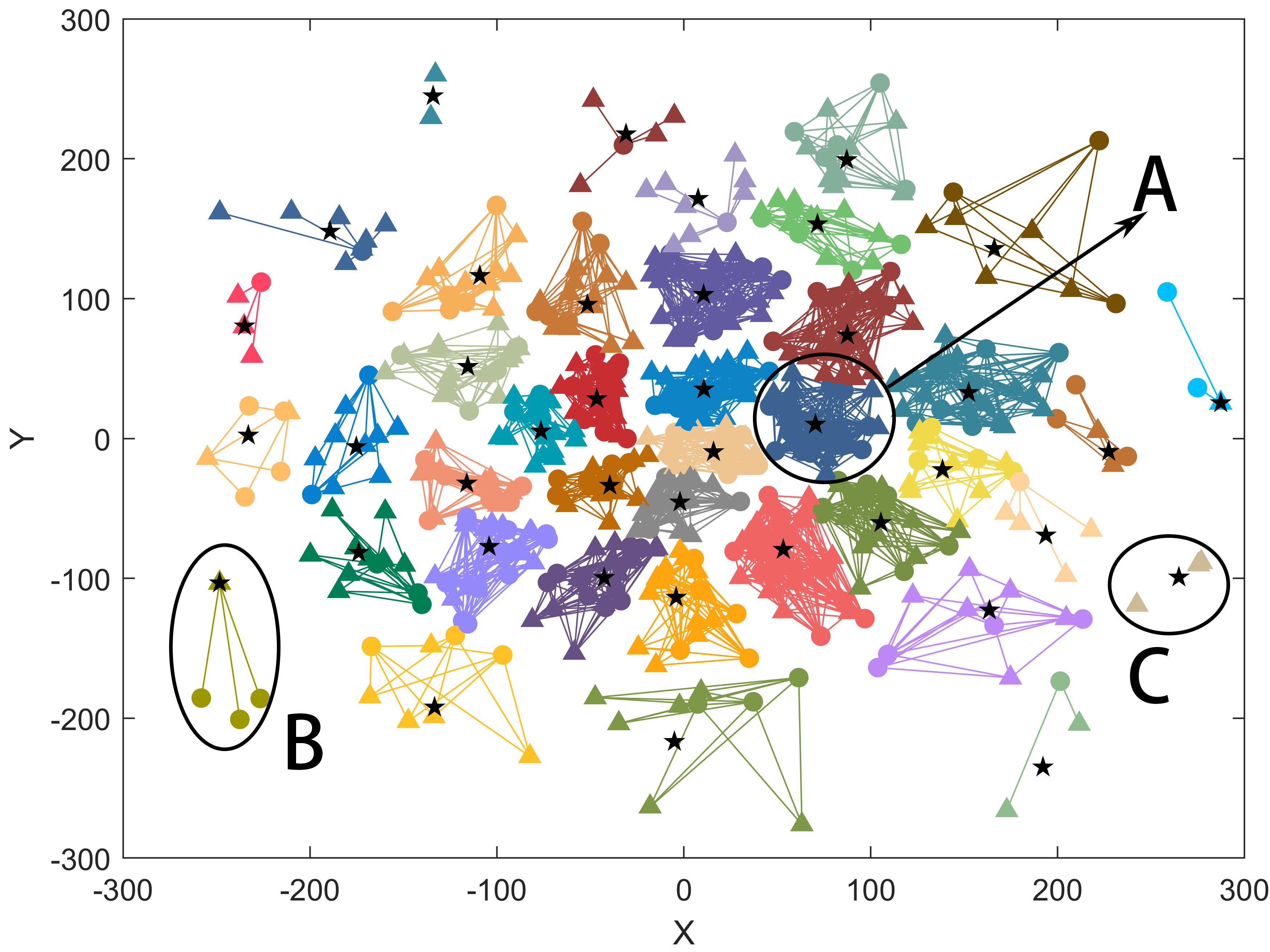}}
		\hspace{0.05\linewidth}
		\subfigure[]{\label{bb}
			\includegraphics[width=0.42\linewidth]{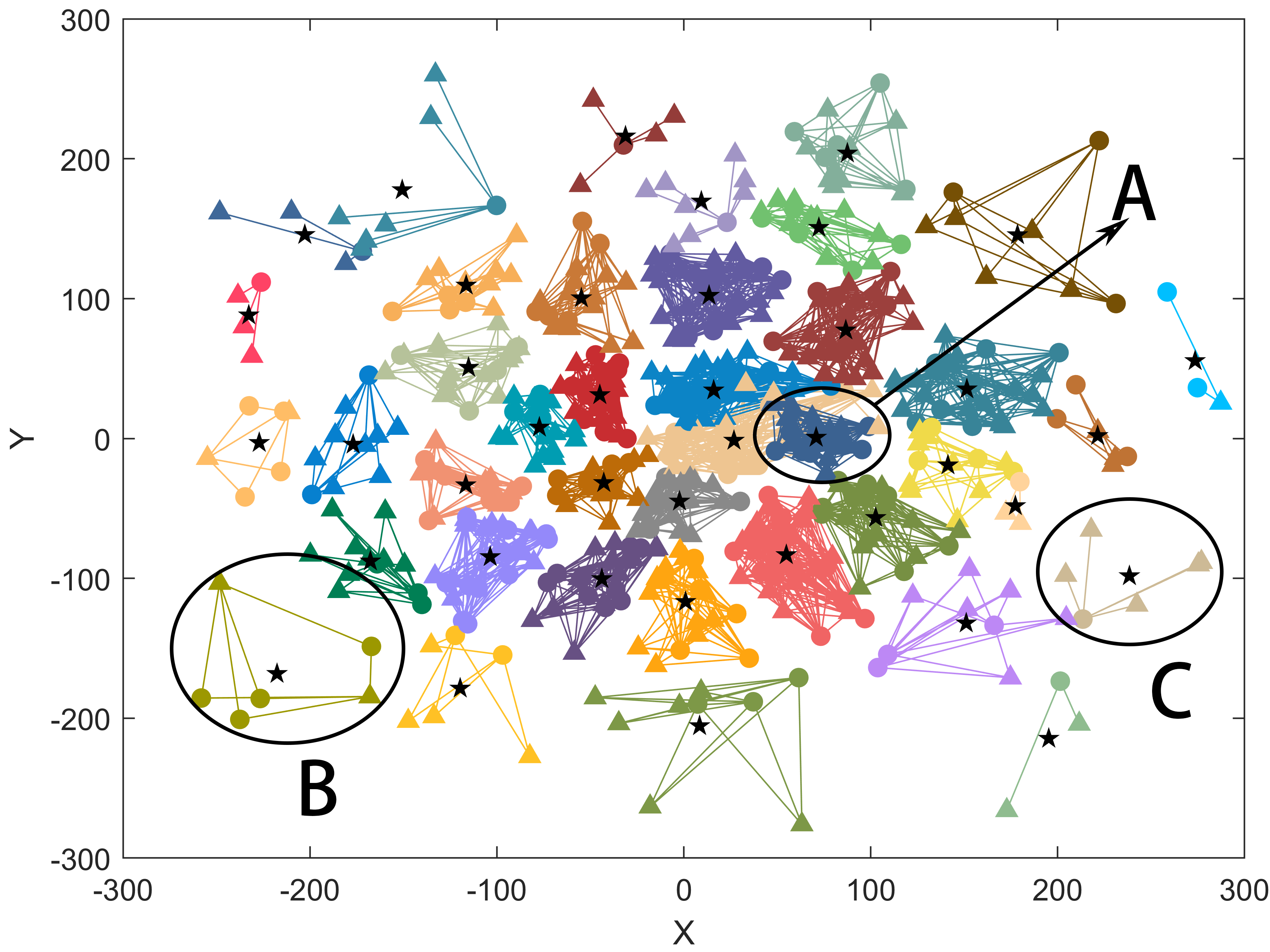}}
		\caption{Visualization of BS-clustering and CGN architectures, with $K<M$. $K = 200, M=400$. Both BSs and users are distributed according to GD. Triangles represent BSs, circles represent users, and stars represent the centroids of clusters. (a) BS-clustering. (b) CGN.}
		\label{The results of simulation of b}
		\vspace{-0.5cm}   
	\end{figure*}
	\vspace{-0.5cm}
	\subsection{Architecture Visualization}\label{ArV}
	We now demonstrate the BS-clustering and the CGN architectures. Simulation settings are shown in Table \ref{Simulation Settings}. Assume both BSs and users are distributed in the network area with the inconstant density\footnote{For visualization, we plot networks with fewer number of nodes in order to clearly show the architectures.}. Here, we choose Gaussian distribution (GD) as an example, to mimic the city with denser network nodes in the central part while sparser network nodes in the marginal area. Specifically, both users and BSs are distributed according to $\mathcal{CN}(0,1)$. BS-clustering architecture is generated by decomposing all the BSs into $L$ non-overlapping clusters according to $k$-means++ algorithm.
	CGN is generated according to our Algorithm \ref{algorithm}. We plot CGN and BS-clustering architectures with $K>M$ in Fig. \ref{The results of simulation of u}. To unfold the generality of our CGN, we plot another case with $K<M$ in Fig. \ref{The results of simulation of b}.
	\begin{table}[htbp]
		\vspace{-0.3cm}
		\caption{Simulation Settings}
		\vspace{-0.3cm}
		\begin{center}
			\begin{tabular}{c|c}
				\hline \bfseries Definition and Symbol & \bfseries Value \\
				\hline
				The number of clusters ($L$) & 40 \\
				Power constraint ($P$) & 1 W\\
				Noise power ($N_{0}$) & 0.09 W\\
				Path loss factor ($\alpha$) & 4\\
				Distance threshold ($d_{0}$) & 5 m\\
				Capacity difference threshold ($\delta$) & 0.2 bps \\   
				\hline
			\end{tabular}
			\label{Simulation Settings}
		\end{center}
	\end{table}
	
	\vspace{-0.5cm}
	In Fig. \ref{The results of simulation of u} and Fig. \ref{The results of simulation of b}, BSs and users are represented by triangles and circles, respectively. A cluster is represented by network nodes interconnected with lines using the same color. Different colored areas correspond to different clusters. It can be observed in Fig. \ref{au} and Fig. \ref{ab} that there are many clusters with BSs only, leading to a waste of resource and energy. Such phenomenon happens in BS-clustering architectures, no matter $K>M$ or $K<M$. In Fig. \ref{bu} and Fig. \ref{bb}, according to the CGN architecture, this phenomenon can be totally avoided, since the CGN is designed based on the cluster sum capacity, which are determined by BSs and users jointly. To give a more in-depth analysis on the differences between BS-clustering and CGN architectures, we select several representative clusters to elaborate:
	\begin{itemize}
		\item[(i)] The clusters with larger sum capacity in the BS-clustering architecture will become clusters with medium sum capacity in the CGN architecture, marked as cluster A in Fig. \ref{The results of simulation of u} and Fig. \ref{The results of simulation of b}. In Fig. \ref{au}, the sum capacity of cluster A is 0.25 bps, while is 0.16 bps in Fig. \ref{bu}, closer to the average cluster sum capacity 0.07 bps. Similarly, the sum capacity of cluster A is 0.26 bps in Fig. \ref{ab} while is 0.2 bps in Fig. \ref{bb}.
		\item[(ii)] On the contrary, some clusters with smaller sum capacity in the BS-clustering architecture will be merged and reorganized into clusters with larger sum capacity in the CGN architecture, such as cluster B in Fig. \ref{The results of simulation of u} and Fig. \ref{The results of simulation of b}. In both Fig. \ref{au} and Fig. \ref{ab}, there is only one BS serving multiple users in cluster B, leading to the sum capacity of cluster B is only $1.6 \times 10^{-6}$ bps in Fig. \ref{au}, and is $7.0 \times 10^{-7}$ bps in Fig. \ref{ab}. In the CGN architecture, cluster B contains more BSs and users. The sum capacity of cluster B becomes $2.2 \times 10^{-4}$ bps in Fig. \ref{bu} and becomes $1.5 \times 10^{-5}$ bps in Fig. \ref{bb}. 
		\item[(iii)] In the BS-clustering architecture, there exist clusters without users, such as cluster C. This is one of the shortcomings of the BS-clustering architecture: it is impossible to prevent the existence of clusters with only one kind of network nodes. The CGN network architecture can perfectly avoid this problem. It makes the sum capacity of all clusters as equal as possible, both BSs and users should exist in all the clusters. Obviously, in Fig. \ref{au} and in Fig. \ref{ab}, the sum capacity of cluster C is 0. In Fig. \ref{bu} and Fig. \ref{bb}, the cluster sum capacity become $3.8 \times 10^{-5}$ bps and $2.4 \times 10^{-5}$ bps, respectively.
	\end{itemize}
	\subsection{Performance Comparison}
	In this subsection, we compare the performance across CGN, BS-clustering and cellular architectures. Unlike the choose of fewer nodes for visualization, now we adopt ultra-dense network nodes for accuracy. In simulation, we consider a square network area with side length $a$. Denote $K = \rho_{u}a^{2}$ and $M = \rho_{b}a^{2}$, where $\rho_{u}$ and $\rho_{b}$ are the density of users and BSs, respectively. Two kinds of distributions of network nodes are considered: the uniform distribution (UD) and the GD.
	
\begin{figure}[htbp]
		\centering
		\hspace{-0.08\linewidth}
		\subfigure[]{
			\includegraphics[width=0.87\linewidth]{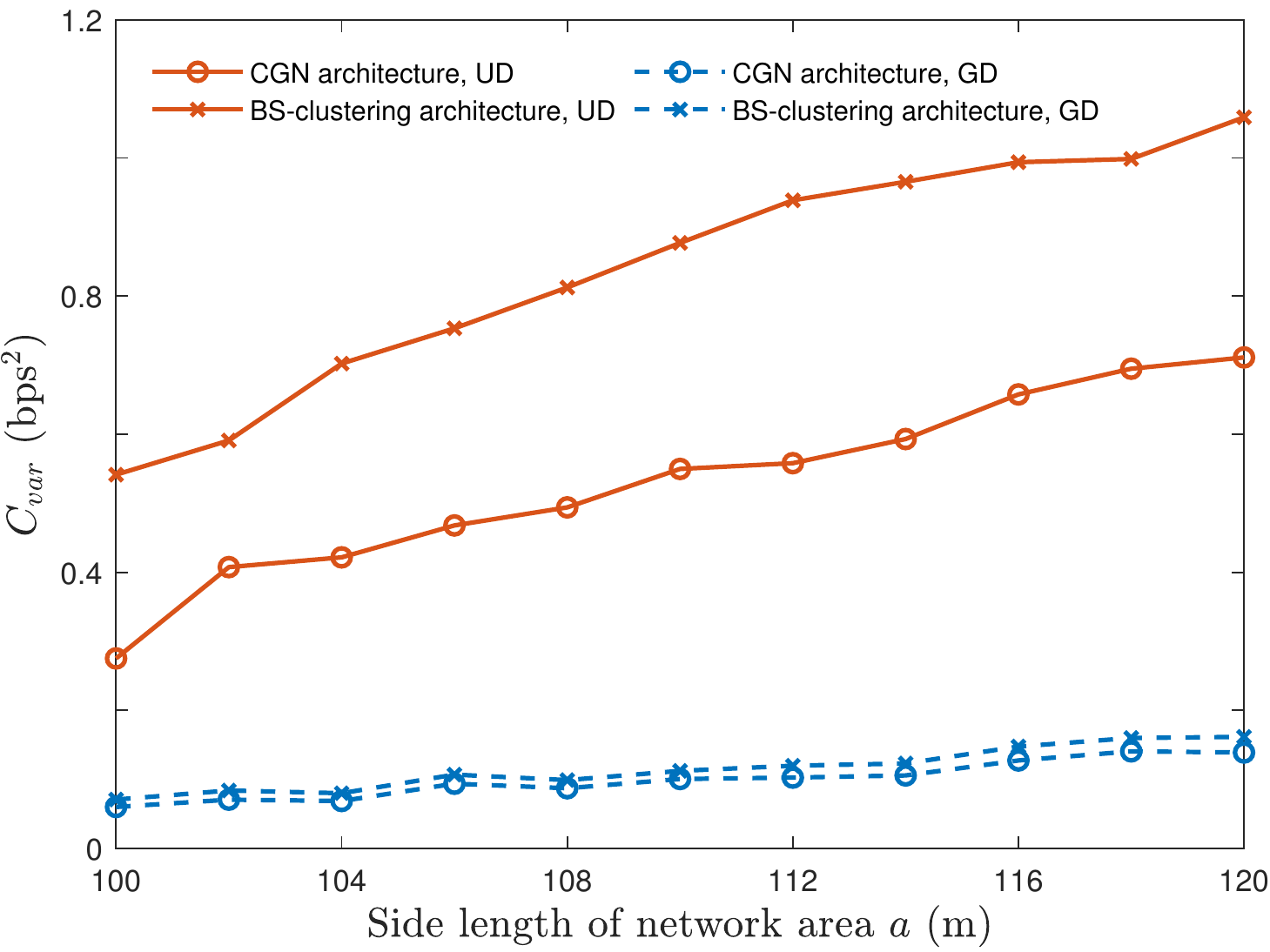}}
		\caption{Comparison between CGN and BS-clustering architectures in terms of $C_{var}$. Two distributions of network nodes are considered: UD and GD.}
		\label{Comparison_var}
		\vspace{-0.5cm}    
	\end{figure}

	The variances of the cluster sum capacity $C_{var}$ of the CGN architecture and the BS-clustering architecture are shown in Fig. \ref{Comparison_var}. It can be observed that $C_{var}$ of the CGN architecture is smaller. Especially in the UD case, $C_{var}$ of CGN decreased by at least 30.4\% compared to $C_{var}$ of BS-clustering. It means the CGN architecture can guarantee a more fair QoS for all the users, which is consistent with our Theorem \ref{thm}.
	
\begin{figure}[htbp]
		\centering
		\hspace{0.13\linewidth}
		\subfigure[]{\label{Comparison_min}
			\includegraphics[width=0.9\linewidth]{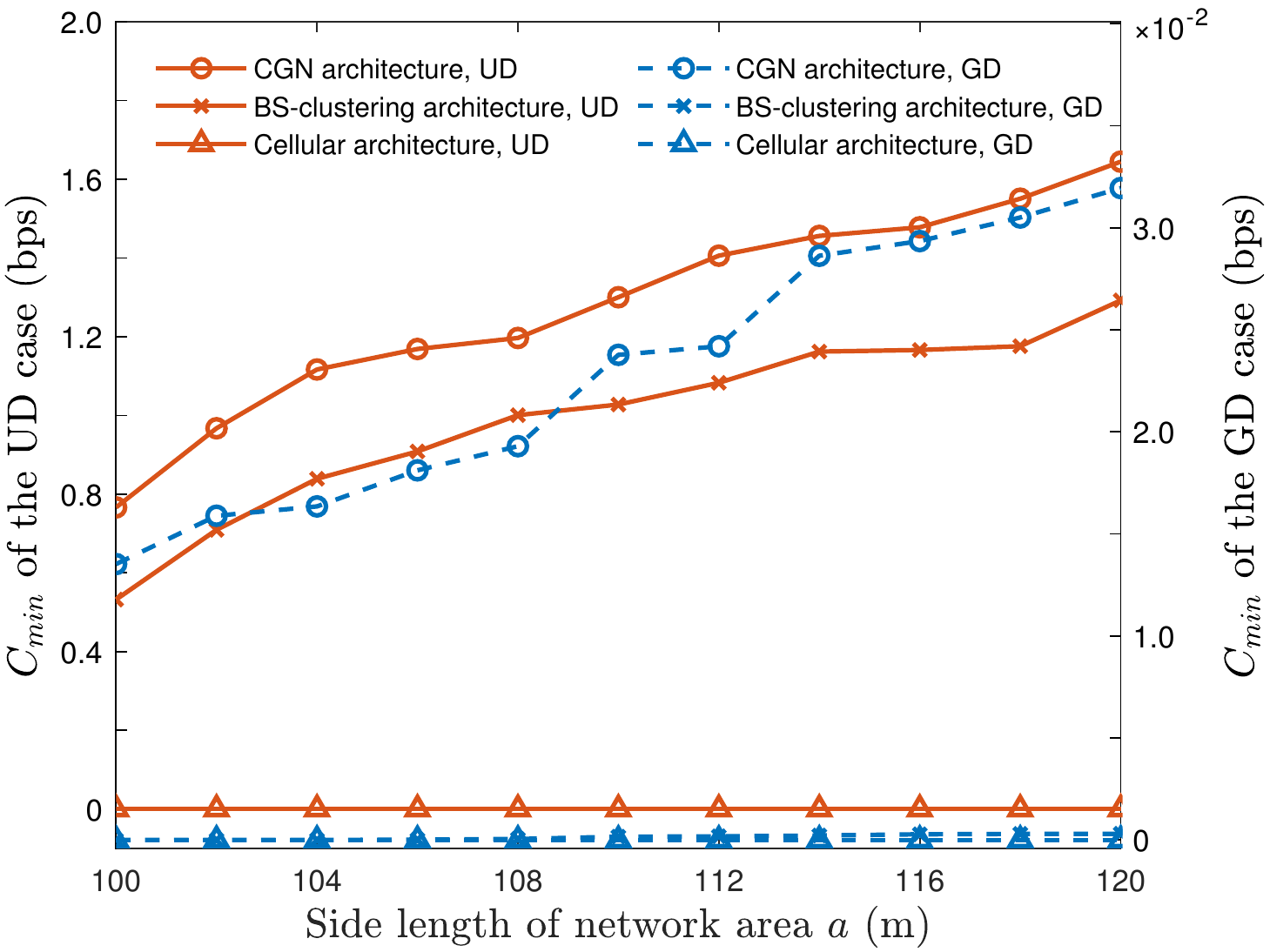}}
		\vfill
		\hspace{-0.03\linewidth}
		\subfigure[]{\label{Comparison_average}
			\includegraphics[width=0.88\linewidth]{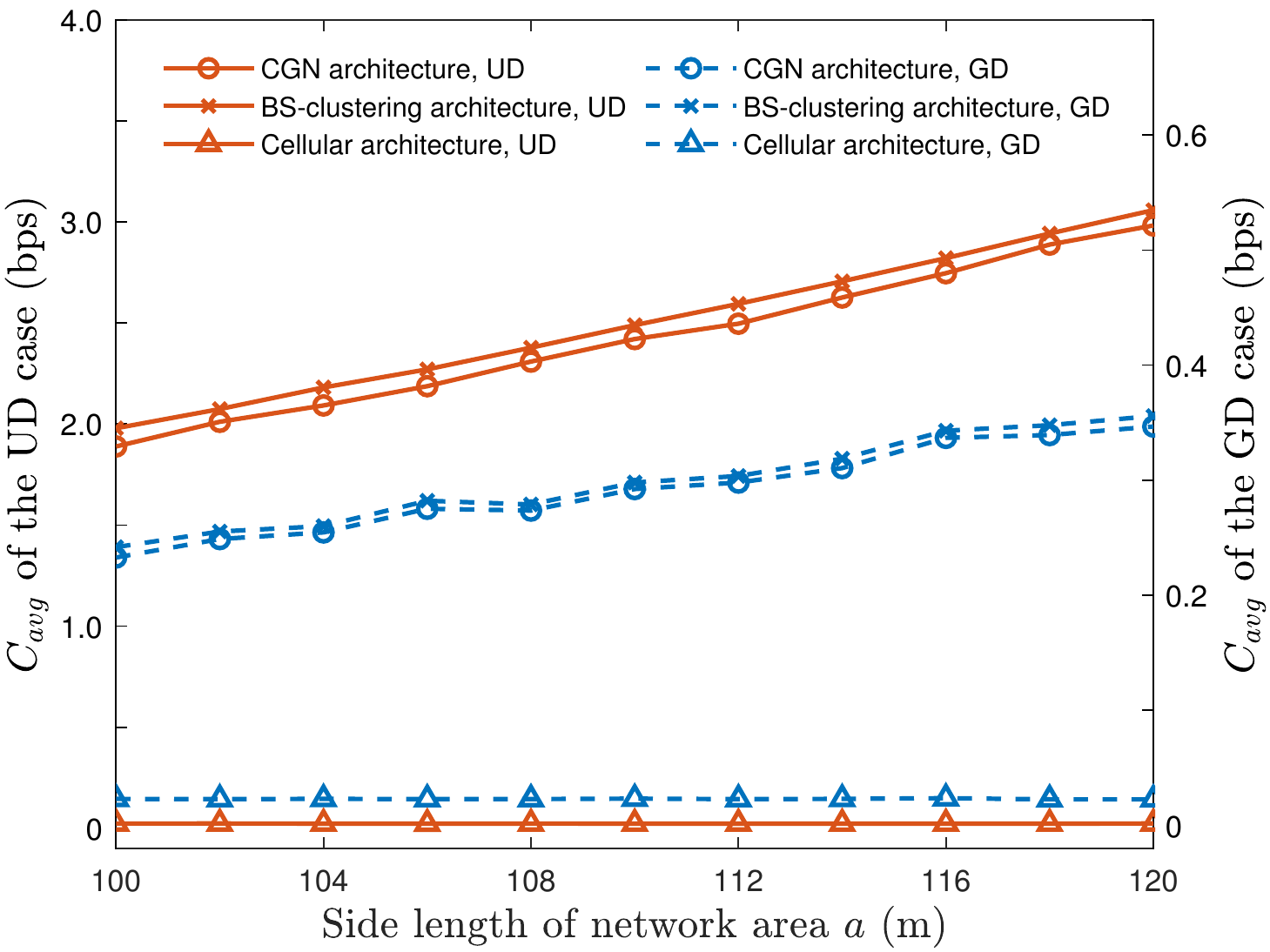}}
		\caption{Comparison across CGN, BS-clustering and cellular architectures in terms of $C_{min}$ and $C_{avg}$. Two distributions of network nodes (BSs and users) are considered: UD and GD. (a) $C_{min}$. (b) $C_{avg}$.}
		\label{Comparison}
		\vspace{-0.5cm}    
	\end{figure}

	The minimum cluster sum capacity $C_{min}$ of three architectures are plotted in Fig. \ref{Comparison_min}. It can be observed that $C_{min}$ of the CGN architecture is the highest, independent of the network areas and the distributions of network nodes. Compared to the BS-clustering architecture, $C_{min}$ of the CGN architecture with UD has at least 30\% performance gain, while with GD, such performance gain can achieve at least 100 times. This is because in the BS-clustering architecture, there always exist clusters with extremely small number of network nodes, leading to the extremely low cluster sum capacity, as illustrated in Fig. \ref{au} and Fig. \ref{ab}. As for the cellular architecture, $C_{min}$ is always zero (i.e., the lowest), corresponding to the cluster containing a single BS without users.
	
	In Fig. \ref{Comparison_average}, we plot the average cluster sum capacity $C_{avg}$ of three different architectures, where $C_{avg} = (\sum_{l=1}^{L}C_{l})/L$. Compared with the cellular architecture, $C_{avg}$ of our CGN architecture has at least 80.3 times performance gain in the UD case, and at least 9.1 times gain in the GD case. Compared with the BS-clustering architecture, $C_{avg}$ in our CGN has a slight reduction of 3.2\% in the UD case, and 2.4\% in the GD case. Combined with Fig. \ref{Comparison_min}, it can be found that these slight degradation on $C_{avg}$, which can be ignored in practice, brings great performance gain on $C_{min}$, such as at least 100 times performance gain in the GD case and 30\% performance gain in the UD case. As a brief summary, our CGN provides great performance gains on both $C_{min}$ and $C_{avg}$ compared with cellular architectures; it can also provide great performance gains on $C_{min}$ compared with BS-centric clustering structures, but with a slight reduction on $C_{avg}$. Generally speaking, the performance of CGN is better than both BS-clustering and cellular architectures.
	\section{Conclusion}
	In this article, we propose a novel CGN architecture for future ultra-dense wireless systems. It guarantees high capacity and high scalability simultaneously. The design essence of CGN is summarized as our Theorem \ref{thm}; the whole network should be decomposed into non-overlapping clusters with equal cluster sum capacity. Simulation results show that our CGN achieves higher capacity, which means the cell/cluster-edge problems in traditional BS-centric architectures can be alleviated greatly. Specifically, in terms of the minimum cluster capacity, our CGN outperforms BS-clustering architecture, with performance gains of at least 30\%. In terms of the average capacity per BS, our CGN outperforms cellular with gains of at least 9.1 times. Our CGN also has superior scalability; the increase of network nodes in a cluster will not affect the signaling overhead and computation complexity of other clusters. More importantly, both of our theorem and CGN architecture are general enough and applicable for various scenarios with different distributions of BSs and users.

	\bibliographystyle{IEEEtran.bst}
	\bibliography{ref/refs.bib}
	
\end{document}